\documentclass[aps,twocolumn,prb,floatfix,superscriptaddress,footinbib,showpacs]{revtex4-1}

\usepackage{color}
\usepackage{verbatim}
\usepackage{hyperref}
\usepackage{epsf}
\usepackage{graphics}
\usepackage{graphicx}
\usepackage{dcolumn}
\usepackage{bm}
\usepackage{amsmath}
\usepackage{color}

\begin{document} 
\title{Semiclassical Klein Tunneling and Valley Hall Effect in Graphene}

\author{Christoph M. Puetter}
\email{cpuetter@comas.frsc.tsukuba.ac.jp}
\affiliation{Graduate School of Pure and Applied Sciences, University of Tsukuba, 1-1-1
  Tennodai, Tsukuba, 305-8577, Japan}
\author{Satoru Konabe}
\affiliation{Graduate School of Pure and Applied Sciences, University of Tsukuba, 1-1-1
  Tennodai, Tsukuba, 305-8577, Japan}
\affiliation{CREST, Japan Science and Technology Agency, 4-1-8 Honcho Kawaguchi,
  332-0012, Japan}
\author{Yasuhiro Tokura}
\affiliation{Graduate School of Pure and Applied Sciences, University
  of Tsukuba, 1-1-1
  Tennodai, Tsukuba, 305-8577, Japan}
\author{Yasuhiro Hatsugai}
\affiliation{Graduate School of Pure and Applied Sciences, University of Tsukuba, 1-1-1
  Tennodai, Tsukuba, 305-8577, Japan}
\affiliation{Center for Integrated Electronic Systems, Tohoku University, Sendai 980-8578, Japan}
\author{Kenji Shiraishi}
\affiliation{Graduate School of Engineering, Nagoya University,
  Nagoya, 464-8603, Japan}
\affiliation{Graduate School of Pure and Applied Sciences, University of Tsukuba, 1-1-1
  Tennodai, Tsukuba, 305-8577, Japan}
\affiliation{Center for Integrated Electronic Systems, Tohoku University, Sendai 980-8578, Japan}

\begin{abstract}
  We study the dynamics of semiclassical electrons in
  (gapped) graphene in two complementary limits, 
  i.e. in the Klein tunneling and valley Hall effect regimes,
  by scattering wave packets off armchair step
  potentials and by exposing wave packets to a uniform electric field,
  respectively.
  Our numerical wave packet simulation
  goes beyond semiclassical analytical approximations and standard Klein
  tunneling treatments and allows to study intra- and intervalley scattering processes.
  We find distinct Klein tunneling characteristics for low and tall steps, which 
  include unusual Berry curvature induced side shifts of the
  scattered wave packet trajectories.
  In the presence of a uniform field, our simulations capture 
  the semiclassical valley Hall effect which manifests in the form of laterally shifted Bloch oscillations.
  Such anomalous trajectory corrections can 
  be relevant for Klein tunneling experiments and electron optics devices.
  We present detailed simulation results.
\end{abstract}
\pacs{72.80.Vp}
\maketitle

\section{Introduction}
Graphene possesses extraordinary electronic transport properties 
due to the two special points in the band structure 
where conduction and valence band meet.
At these so-called Dirac points the energy dispersion is conical,
and electrons circling around each Dirac point
acquire a nontrivial Berry phase of $\pm \pi$. \cite{CastroNeto09RMP}
A nontrivial Berry phase can lead to
remarkable transport phenomena as displayed by  
the half-integer quantum Hall effect in graphene
\cite{Novosolov05Nature,Zhang05Nature,Hatsugai06PRB}
and the intrinsic anomalous Hall effect in ferromagnetic materials.
\cite{Nagaosa10RMP}
The valley Hall effect and the Klein paradox, too, are closely associated with
the conical Dirac dispersion. 
In gapped graphene and related materials such as silicene and
germanene, \cite{Takeda94PRB,Lalmi10APL,Padova10APL,Liu11PRB,Ezawa13PRL}
both the valley Hall effect and the Klein paradox 
can provide striking signatures of the underlying nontrivial band
topologies and, in general, involve both Dirac valleys.

The Klein paradox refers to the almost certain transmission of
relativistic Dirac fermions across a potential barrier
regardless of its height and width. \cite{Klein29ZPhys}
Considerable effort has recently been put into 
observing and exploring the Klein phenomenon in graphene devices,
\cite{Williams07Science,Huard07PRL,Gorbachev08NanoLett,Stander09PRL,Young09NatPhys,Sajjad12PRB,Reijnders13AnnPhys,Rahman13arXiv}
with evidence reported based on resistance measurements across a 
potential barrier and comparison with theoretical predictions
for ballistic and diffusive transport; \cite{Stander09PRL}
conductance oscillations across a potential barrier; \cite{Young09NatPhys}
and characteristic angle-selective transmission at pn junctions.
\cite{Sajjad12PRB,Rahman13arXiv}
In contrast, the valley Hall effect occurs in the opposite limit where 
a sharp step potential or barrier potential is replaced by a uniform potential gradient,
leading to electronic Bloch oscillations in ideal dissipationless systems.
Valley-based phenomena have recently attracted particular interest
due to the technological potential of spin-valley selective
photoelectric excitations in the quasi-2D transition metal dichalcogenides. 
\cite{Mak10PRL,Xiao12PRL,Mak12NatNano,Cao12NatComm,Mak14Science}

In the present paper we utilize a semiclassical wave packet simulation.
A semiclassical approximation can provide 
unique insight into the effective dynamics of Bloch electrons 
in bands with nontrivial topologies.
\cite{Chang08JPhysCondensMatter,Xiao10RMP}
This is evident from the semiclassical equations of motion
\begin{eqnarray}
  \label{eq:SemiClassEQM}
  \dot{\bf r} = \nabla_{\bf k} \epsilon_{\bf k} - \dot{\bf k}
  \times \hat{\bf \Omega}_{\bf k}, 
  \hspace{0.5cm}
  \dot{\bf k} = -e ({\bf E} + \dot{\bf r} \times {\bf B}),
\end{eqnarray} 
where ${\bf r}$ stands for the electron position, $\epsilon_{\bf k}$
for the band energy at crystal momentum ${\bf k}$, 
$e$ for the elementary charge, ${\bf E}$
and ${\bf B}$ for electric and magnetic field, respectively,
while, crucially, the first cross product term 
represents a correction due to the
Berry curvature ${\bf \Omega}_{\bf k}$ and the driving force $\dot{\bf
  k}$. 
A wave packet simulation has further advantages.
It goes beyond the standard analytical
treatment of Klein tunneling in graphene, which usually relies 
on linearized single valley Dirac bands
and a plane wave ansatz, and hence is insufficient to 
capture intervalley scattering and Berry phase related effects.
\cite{Katsnelson06NatPhys,Cheianov06PRB,Peres09JPhysCondensMatter,Allain11EurPhysJB}
In addition, in contrast to the semiclassical analytical 
formalism [exemplified by Eq. (\ref{eq:SemiClassEQM})], 
which requires smooth fields, single or degenerate
bands and uniquely
defined wave packet centres in real space and in momentum space,
\cite{Xiao10RMP,Gao14arXiv}
a wave packet simulation is in principle exact and generalizes 
the semiclassical approach to arbitrary potentials and multi-band systems.

\begin{figure}[t!]
  \centering
  \includegraphics*[width=1.0\linewidth, clip]{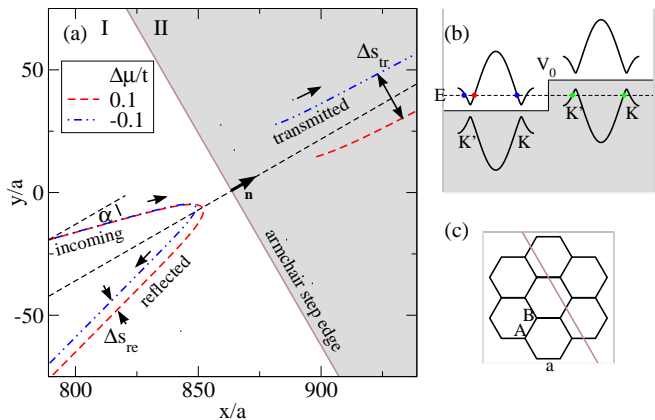}
  \caption{(Color online) 
    (a) Wave packet trajectories for different staggered sublattice potentials 
    ($\Delta \mu = \pm 0.1 t$) yielding different Berry
    curvatures. 
    The step edge and step normal are indicated by the solid brown and
    dashed black lines, respectively, while
    region I (II) mark the lower (upper) portions of the step
    where $V({\bf r}) = 0$ ($V_{0}$).
    Here the step is sharp,  i.e. $\Delta w = 0$.
    The trajectories near the step edge are in general not well defined and have
    been omitted for the transmitted wave packet portions for clarity.
    (b) Schematic of the Klein tunneling energy-momentum relation
    at an armchair potential step. 
    (c) Orientation of the armchair edge (brown line) on the graphene
    lattice.  
    See main text for details.
    \label{fig:trajectories}}
\end{figure}

For simplicity we focus on the dynamics of semiclassical 
spinless electrons in the Klein tunneling and valley Hall effect
regimes, by scattering wave packets off armchair 
potential steps and by imposing a smooth potential gradient, respectively, 
while taking the full graphene tight-binding Hamiltonian into account.
Our main findings are two-fold.
First, our numerical study
reveals distinct Berry curvature induced side shifts in the wave packet motion 
near potential steps [see Fig. \ref{fig:trajectories} (a)] 
that are direct, two-dimensional analogues of those
appearing in the optical Hall effect. \cite{Onoda04PRL,Onoda06PRB}
These shifts occur even in the sharp step limit where 
the driving force is almost singularly confined to the 
step edge and indicate deviations from simple reflection and
refraction laws.
Since we account for the full dispersion,
the transmission and reflection probabilities, in addition to the
lateral shifts, vary with height and width of the potential step.
This behavior is a consequence of distinct intra- and intervalley
scattering regimes and give rise  to substantial corrections to 
the usual Klein tunneling process for relativistic Dirac particles.
\cite{Klein29ZPhys,Peres09JPhysCondensMatter,Allain11EurPhysJB}

\begin{figure}[t!]
  \centering
  \includegraphics*[width=1.0\linewidth, clip]{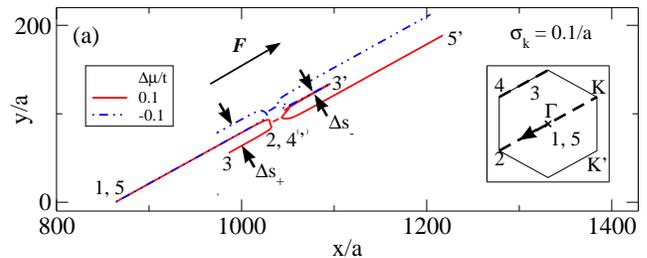}
  \caption{(Color online) 
    Bloch oscillations / valley Hall effect of wave packets in the
    presence of a uniform electric field with magnitude $F  = 0.015 \,
    t/(|e| \, a)$.
    The main panel shows the 
    real space wave packet trajectories of electron and hole band 
    portions for different staggered sublattice
    potentials ($\Delta \mu = \pm 0.1 \, t$).
    The inset displays the corresponding trajectory in momentum space.
    The numbers in the main panel and in the inset correspond to each
    other and indicate temporal order.
    See main text for details.
    \label{fig:BO}}
\end{figure}

Second, in the opposite case of an extended linear potential, we find
Bloch-Zener oscillations that acquire a finite component perpendicular to the
potential gradient (see Fig. \ref{fig:BO}). 
These modified oscillations are a manifestation of the valley Hall effect where 
the trajectory side shifts are induced by a
finite Berry curvature near the Dirac points. \cite{Diener03arXiv,Xiao10RMP}
The present simulations allow to investigate in detail the accompanying 
interband Bloch-Zener transitions 
that occur near the Dirac points and are closely related to the 
transmitted wave packet portions in the Klein tunneling regime.
The interband transitions lead to two wave packet portions with
opposite velocities and trajectories as indicated
e.g. by the two red solid lines in Fig. \ref{fig:BO}.

The paper is organized as follows. 
In Sec. \ref{sec:II} we introduce the model, the simulation details and
useful definitions.
In Sec. \ref{sec:III} we present the results for 
semiclassical Klein tunneling at sharp and smooth potential
steps using the full graphene tight-binding Hamiltonian.
To illuminate the basic intra- and inter-valley scattering processes, 
the first subsection (\ref{subsec:IIIA}) discusses in detail
the transmission and reflection probabilities and 
their valley affiliations. 
The second subsection (\ref{subsec:IIIB}) 
focusses on the lateral trajectory shifts induced near a step
interface by a finite Berry curvature.
Sec. \ref{sec:IV} is devoted to the semiclassical valley Hall effect.
Here we apply a uniform potential gradient and present the 
simulation results for the wave packet Bloch oscillations and the
accompanying trajectory side shifts.  
The last section closes with a summary and a brief discussion of
the findings.

\section{Semiclassical wave packet propagation}
\label{sec:II}

To study the semiclassical single electron dynamics we 
consider a spinless electron wave packet moving in 
a two-dimensional honeycomb lattice towards a potential step.
Although it is experimentally challenging to fabricate
graphene devices with thin junction interfaces, \cite{Huard07PRL}
a narrow potential step is the simplest setup 
and an instructive limit to investigate
semiclassical Klein tunneling on a lattice.
In addition, narrow potential steps may also arise accidentally from the presence
of impurities or imperfections of the underlying substrate.
We contrast the Klein tunneling scattering dynamics at a
potential step with the Bloch-Zener oscillation 
dynamics in the presence of an extended linear potential, 
which can be imposed by applying a bias voltage. 
A staggered sublattice potential, which opens an energy gap 
and leads to a finite Berry curvature
near the Dirac points, may be induced by an adequate substrate, or, in the
case of related buckled quasi-two-dimensional materials like silicene or
germanene, by applying a perpendicular electric field.
In the latter materials, however, 
a splitting of the Bloch bands due to significant spin-orbit interaction 
needs to be taken into account. \cite{Liu11PRB}

\begin{figure*}[t!]
  \centering
  \includegraphics*[width=1.0\linewidth, clip]{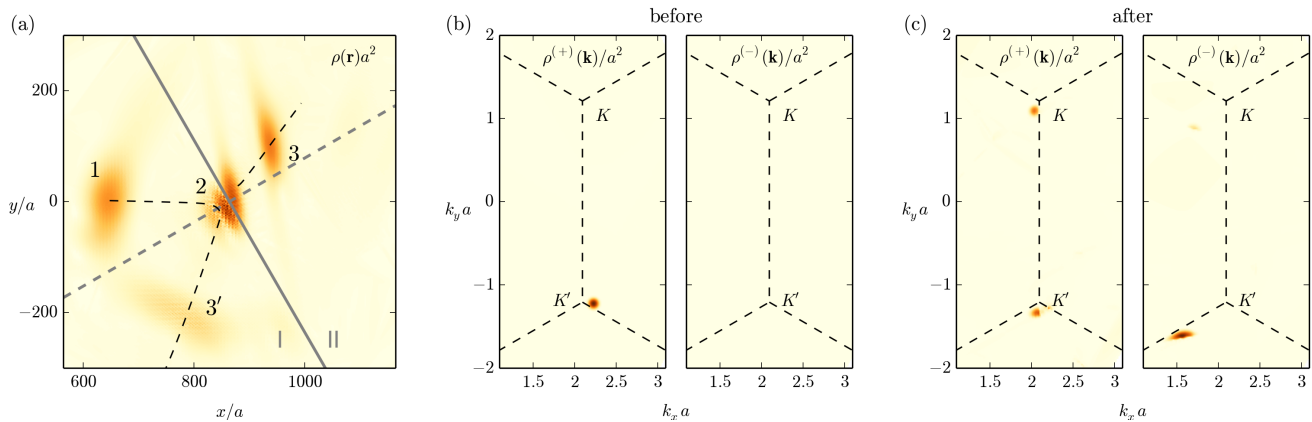}
  \caption{(Color online) 
    (a) Real space snapshot sequence of a wave packet scattering at an
    armchair step potential (indicated by the grey solid line) before (1),
    at (2) and after scattering (3, 3').
    After scattering the wave function consists of 
    transmitted (3) and reflected (3') portions.
    The grey dashed line is normal to the step edge, while the black
    dashed lines indicate the centre of mass trajectories in regions I
    and II.
    (b) Initial momentum space snapshot corresponding to 
    instance 1 in panel (a).
    Only electron states near K' contribute to the wave packet
    composition.
    (c) Momentum space snapshot after scattering representing instance \{3, 3'\} [see
    panel (a)]. 
    The transmitted (3) portion's
    major (minor) contribution stems from the hole band near K' (K),
    while the reflected (3') wave packet portion has contributions
    from electron states near both K and K'.
    See main text for details.
    \label{fig:snapshots}}
\end{figure*}

In the simulation the time evolution of the wave 
packet $| \psi(\tau) \rangle$ is governed
by the time-dependent Schroedinger equation
$i \partial_{\tau} | \psi(\tau) \rangle = (H_{0} + H_{V}) | \psi(\tau) \rangle$,
which contains the usual honeycomb tight-binding
Hamiltonian  \cite{CastroNeto09RMP}  
\begin{eqnarray}
  H_{0} &=& -t \sum_{{\bf r}} (c_{{\bf r}}^{a \dagger} c_{{\bf r}}^{b} 
  + c_{{\bf r}}^{a \dagger} c_{{\bf r}-{a}_{1}}^{b} + c_{{\bf r}}^{a \dagger}
  c_{{\bf r}-{a}_{1}+{a_{2}}}^{b} + \text{h. c.}) \nonumber \\
  &&  + \Delta \mu \sum_{{\bf r}} (c_{{\bf r}}^{a \dagger} c_{{\bf
      r}}^{a} - c_{{\bf r}}^{b \dagger} c_{{\bf r}}^{b} )
\end{eqnarray}
and the external potential 
\begin{eqnarray}
  H_{V} &=&
  \sum_{{\bf r}} [ V({\bf r}) c^{a \dagger}_{{\bf r}} c^{a}_{{\bf r}} +
  V({\bf r}+\vec{\bf \delta}) c^{b \dagger}_{{\bf r}} c^{b}_{{\bf r}}].
\end{eqnarray}
Here $t$ denotes the nearest-neighbor hopping integral, ${\bf a}_{1}$
and ${\bf a}_{2}$ the lattice vectors and $\Delta \mu$ the staggered sublattice potential.
In the following, 
the external potential $V({\bf r}) = V_{0} \; g[{\bf n} \cdot ({\bf r}-{\bf
  r}_{0})]$ either represents a potential step 
of height $V_{0}$, location ${\bf r}_{0}$ and step normal direction
${\bf n}$ [see Fig. \ref{fig:trajectories} (a)] 
or a linear electric potential $V({\bf r}) = - e \; {\bf F} \cdot {\bf
  r}$ with field ${\bf F}$ and electron charge $e$ ($<0$).
The potential step profile is given by 
\begin{eqnarray}
  g(x) &=& 
  \begin{cases}
    0, & x < - \frac{\Delta w}{2} \\
    \Big(\frac{1}{2} + \frac{x}{\Delta w} \big), &  - \frac{\Delta w}{2} \leq  x \leq \frac{\Delta w}{2}\\
    1, & \frac{\Delta w}{2} < x
  \end{cases},
\end{eqnarray}
where $\Delta w$ denotes the step width
[if $\Delta w = 0$ then $g(x) = a \, \delta(x)$, where $a$ is the
nearest-neighbor distance],
The operator $c^{a \dagger}_{{\bf r}}$ ($c^{b \dagger}_{{\bf r}}$)
creates an electron on sublattice
A (B) in unit cell ${\bf r}$, while $\vec{\bf \delta}$ denotes the intra-unit cell 
nearest-neighbor vector.

The wave packet evolution then follows straightforwardly 
from a fourth order split-operator (Suzuki-Trotter) approximation scheme 
$| \psi (\tau) \rangle \approx \Pi_{n=1}^{N} \text{e}^{-i H \Delta
  \tau} |\psi(\tau=0) \rangle$,
where $\Delta \tau = \tau/N$ and $N$ is the total number of iterations.
\cite{Suzuki90PhysLettA,Suzuki92JPSJ}
For simplicity, we impose periodic boundary conditions
and use sufficiently large system sizes and small enough time steps 
to reduce finite size effects.
Although we evolve the wave packet numerically using the real space 
representation, i.e.
\begin{equation}
  | \psi (\tau) \rangle 
  = \sum_{{\bf r}} [\Psi_{A}({\bf r}; \tau) c^{a \dagger}_{{\bf r}}
  + \Psi_{B}({\bf r}; \tau) c^{b \dagger}_{{\bf r}}] |0\rangle,
\end{equation}
where $\Psi_{A (B)}({\bf r}; \tau)$ stands for the wave function on
sublattice A (B),
it is convenient to specify the initial wave packet at $\tau=0$ 
using the eigenbasis of the $H_{0}$.
Thus in the following 
the initial wave packet is given as a Gaussian-like superposition of positive energy
eigenstates centered close to the K' Dirac point 
[see Fig. \ref{fig:snapshots} (b)]
\begin{equation}
  | \psi (\tau=0) \rangle 
  = \sum_{{\bf k}} \Psi^{(+)}_{{\bf k}_{0}, \sigma_{k}, {\bf r}_{0}}({\bf
    k}) \xi^{(+) \dagger}_{\bf k} | 0 \rangle,
\end{equation}
where $\xi^{(+) \dagger}_{{\bf k}}$ creates a state with momentum
${\bf k}$ in the electron band. 
The momentum distribution  
$\Psi^{(+)}_{{\bf k}_{0}, \sigma_{k}, {\bf r}_{0}} ({\bf k})$ 
is centered at ${\bf k}_{0}$ (which, if not stated otherwise, lies close to K') and characterized 
by a width $\sigma_{\text{k}}$, while the initial position in real space is
determined through ${\bf r}_{0}$. \cite{footnote01}
Initializing the wave packet using states from both electron and hole bands
would complicate the propagation dynamics as additional interference 
phenomena such as trembling motion (\emph{Zitterbewegung})
may appear. \cite{Katsnelson07EPJB,Rusin07PRB,Cserti10PRB}

Significant dispersion of free wave packets with central momenta
near the Dirac points \cite{Maksimova08PRB} hampers the
identification of transmitted and reflected trajectory shifts close to
the step edge.
To prevent such a complication, we choose 
the phase of $\Psi^{(+)}_{{\bf k}_{0}, \sigma_{k}, {\bf r}_{0}} ({\bf k})$
such that the real space wave packet 
is most compact and symmetric
at the point of impact onto the step edge
(see Fig. \ref{fig:snapshots}), which is roughly given by ${\bf
  r}_{0}$.
The precise definition of $\Psi^{(+)}_{{\bf k}_{0}, \sigma_{k}, {\bf r}_{0}}
({\bf k})$ is provided in Ref. \onlinecite{footnote01}.

\section{Semiclassical Klein tunneling at an armchair step edge}
\label{sec:III}
In this part we investigate the scattering of wave packets at
sharp and smooth armchair steps with variable step heights, step widths
and wave packet sizes.
For an incoming wave packet with a central momentum close to the K' valley,
the armchair edge is distinct from other edge orientations 
as scattering in general involves states from both K and K' points.
A consequence is that the transmission (reflection) probability does not increase
(decrease) monotonously with the step height as in the standard Klein tunneling
problem with a single, strictly linear Dirac valley. \cite{Allain11EurPhysJB}
In addition, opening a gap by a staggered sublattice potential 
induces a finite Berry curvature that, as shown below, can affect 
outgoing wave packet trajectories and breaks pseudospin
conservation.

The typical real and momentum space evolution of a wave packet scattering off an
armchair step potential is shown in Fig. \ref{fig:snapshots}.
Panel (a) displays the real space probability distribution of the 
wave packet 
$\rho({\bf r}; \tau) = |\Psi_{A}({\bf r}; \tau)|^{2}
+ |\Psi_{B}({\bf r}; \tau)|^{2}$
at three time instances $\tau = 0, 160 \, t^{-1}$ and $300 \, t^{-1}$; 
in addition, panels (b) and (c) indicate the corresponding 
momentum space distributions
$\rho^{(\pm)}({\bf k}; \tau) = |\Psi^{(\pm)}({\bf k}; \tau)|^{2}$
in electron and hole bands before ($\tau = 0$) and after ($ 300 \,
t^{-1}$) scattering, respectively.
The wave functions $\Psi_{A/B}({\bf r}; \tau)$ and
$\Psi^{(\pm)}({\bf k}; \tau)$
are related by a unitary transformation that diagonalizes 
$H_{0}$.
The step height, the step width, the energy and the initial
momentum space size of the wave packet are $V_{0} = 2.0 \, t$,  
$\Delta w = 0$, $E = 0.2 \, t$ and $\sigma_{k} = 0.1/a$, respectively.
The states contributing to the initial wave packet are located near 
the K' valley and have positive energy [see left-hand side of panel (b)].
Upon impact onto the step edge, 
the wave packet splits and is partially transmitted and partially reflected.
The Klein tunneling process involves an interband transition 
from the electron band to the hole band, which causes the appearance of a
nonzero weight on the right-hand side in panel (c).
Furthermore, the left-hand side of panel (c) clearly indicates that,
for the present parameters, 
the reflected wave packet portion consists of contributions from both
Dirac valleys.
Such behaviour is allowed by energy and momentum conservation
and is expected to occur at narrow armchair steps.
More generally, both transmitted and reflected wave packet portions 
can have contributions from K and K' as indicated in
Fig. \ref{fig:trajectories} (b), 
where the energy-momentum of the  incoming wave packet
is indicated in red and the outgoing
energy-momenta are marked in green (transmitted portion) and blue
(reflected portion).

\subsection{Intra- and intervalley scattering}
\label{subsec:IIIA}

\begin{figure}[t!]
  \centering
  \includegraphics*[width=1.0\linewidth, clip]{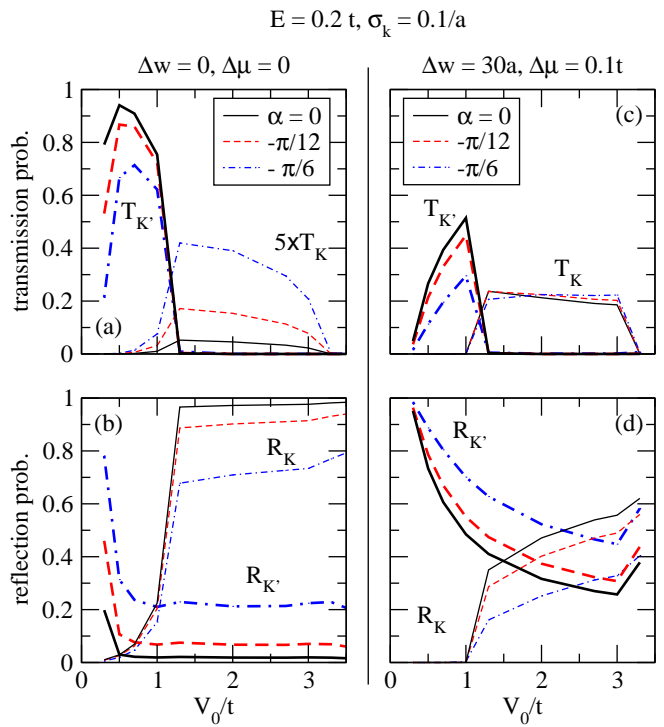}
  \caption{(Color online) 
    Valley-resolved transmission ($T_{K}$, $T_{K'}$) and reflection
    ($R_{K}$, $R_{K'}$)
    probabilities
    for an incoming wave packet of energy $E$ and
    size $\sigma_{k}$ scattering at an armchair step edge
    as a function of step height $V_{0}$.
    Thick (thin) lines represent the K' (K) valley.
    The parameter $\alpha$ denotes the angle of the incoming trajectory 
    relative to the step normal [see Fig. \ref{fig:trajectories} (a)].
    On the left-hand side the step width $\Delta w$ vanishes and
    the bands are gapless ($\Delta \mu= 0$). 
    On the right-hand side the step width is finite but narrow and the band
    structure exhibits a finite gap.
    In (a) the curves for $T_{K}$ have been multiplied by factor of
    $5$ for better visibility. 
    \label{fig:TAndRProbs}}
\end{figure}

Before turning to the lateral shifts associated with reflection and
transmission at the step edge, it is useful to consider the 
reflection and transmission probabilities 
of the Klein tunneling process in more detail.
These are shown in the Fig. \ref{fig:TAndRProbs},
where panels (a) and (b) display the valley-resolved 
transmission and reflection probabilities, respectively,
for a sharp step ($\Delta w = 0$)
in the absence of a staggered sublattice potential ($\Delta  \mu= 0$).
Since for sufficiently large steps ($V_{0} > E$), the transmitted
and the reflected wave packet portions belong to the 
hole and the electron band, respectively,
the valley-resolved transmission and reflection probabilities are
simply defined by
\begin{eqnarray}
  T_{\nu} = \sum_{{\bf k} \in \text{BZ}_{\nu}} \rho^{(-)}({\bf k}; \tau'), 
  \; 
  R_{\nu} = \sum_{{\bf k} \in \text{BZ}_{\nu}} \rho^{(+)}({\bf k}; \tau'),
\end{eqnarray}
where $\nu = K, K'$ and the set $\text{BZ}_{K}$ contains all momenta 
of the first Brillouin zone that are
closer to K than K' and vice versa for $\text{BZ}_{K'}$.
The conservation of probability implies $T_{K} + T_{K'} + R_{K}
+ R_{K'} = 1$.
The time $\tau'$ is chosen sufficiently large so that transmitted and 
reflected real space wave packet portions are well removed from the step edge.

Clearly, transmission and reflection are drastically different for low and high
step heights $V_{0}$ and vary with incoming 
angle $\alpha$ [see Fig. \ref{fig:trajectories} (a)].
At low step heights $V_{0} \lesssim V_{c}$, where $V_{c} = t + E = 1.2
\, t$, the scattering of the
wave packets in Fig. \ref{fig:TAndRProbs} 
is dominated by the K' valley where the initial 
wave packet momentum is located.
At larger steps $V_{0} \gtrsim V_{c}$, intervalley scattering dominates.
Particularly notable is the effect of pseudospin conservation,
\cite{Allain11EurPhysJB} 
which suppresses transmission for $V_{0} \gtrsim V_{c}$
and therefore enhances intervalley reflection.
For step heights greater than $V_{0} \gtrsim 3 \, t + E \; (= 3.2 \, t)$,
transmission is completely suppressed since
no states are energetically  accessible in region II.
The suppression of $T_{K'}$ at $V_{0} \approx 1.2 \, t$ has a similar
origin as the K' valley cannot provide states with suitable 
momenta and energy for transmission [see also Fig. \ref{fig:trajectories} (b)].

The panels (c) and (d) of Fig. \ref{fig:TAndRProbs} display the effect 
of a finite gap $\Delta \mu \neq 0$ and a smooth ($\Delta w \neq 0$) potential step.
Qualitatively, the dependencies of transmission and reflection
probabilities on $V_{0}$ and $\alpha$ are
similar to the $\Delta \mu = 0$ and $\Delta w = 0$ case.
However, the main effect of widening the step and/or inducing a gap is the
enhanced (reduced) transmission at high (low) step heights, i.e. 
increased $T_{K}$ (decreased $T_{K'}$).
This observation even applies when
varying $\Delta \mu$ and $\Delta w$ independently and is consistent
with the lack of pseudospin conservation for finite $\Delta \mu$ 
and $\Delta w$.
Due to the narrow step width, the Klein tunneling process
here can be considered as a nonadiabatic Landau-Zener interband transition,
where the transition probability usually depends sensitively on the transition time
(given by the potential gradient) and the gap size.
 \cite{Zener32ProcRSocLondA,Wittig05JPhysChemB,Fuchs12PRA}

\subsection{Berry curvature induced lateral shifts}
\label{subsec:IIIB}
Since we are concerned with a single step only, we define the
trajectories on either side of the step profile via
\begin{eqnarray}
  \label{eq:trajectories}
  {\bf r}_{M} = \sum_{{\bf r} \in M} {\bf r} \,
  |\Psi_{A}({\bf r}; \tau)|^{2} 
    + \sum_{{\bf r}+\vec{\bf \delta} \in M}({\bf r}+\vec{\bf \delta}) \,  |\Psi_{B}({\bf r}; \tau)|^{2}, 
\end{eqnarray}
where $M = \text{I, II}$ and the regions I and II 
are indicated in Fig. \ref{fig:trajectories} (a).
This trajectory definition is only well defined when the wave function portions on
either side are sufficiently removed from the step interface.
For this reason the transmitted wave packet trajectories in 
Fig. \ref{fig:trajectories} (a) in region II are partially omitted for clarity.
To compute the Berry curvature induced 
lateral shifts, we utilize only the trajectory sections far away from
the step interface.

\begin{figure}[t!]
  \centering
  \includegraphics*[width=1.0\linewidth, clip]{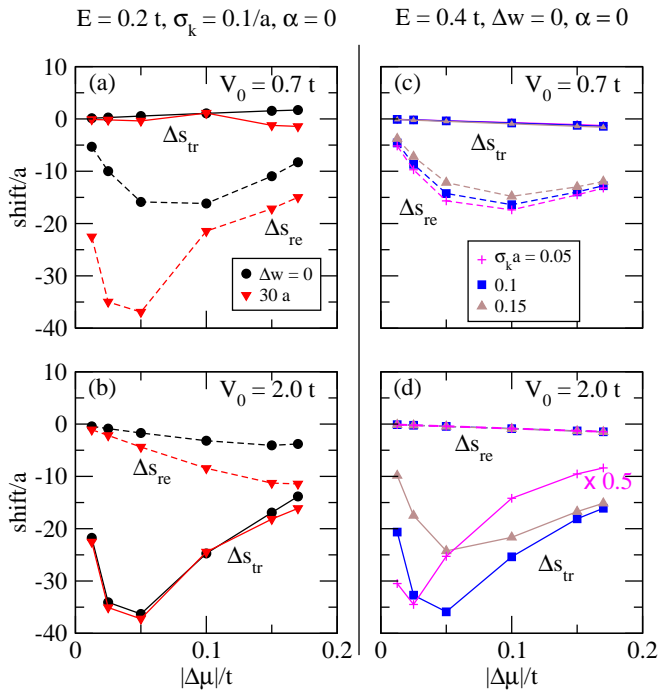}
  \caption{(Color online) 
    Lateral shifts of the transmitted (solid lines) and the reflected (dashed lines) wave packet
    trajectories for normal incidence ($\alpha = 0$) as a function of
    staggered sublattice potential $\Delta \mu$.
    The panels (a) and (b) display the shifts of wave
    packets with energy $E = 0.2 \, t$ ($V_{c} = 1.2 \, t$) for step heights
    $V_{0} = 0.7 \, t$ and $2.0 \, t$, respectively, for two 
    step widths ($\Delta w = 0, 30 \, a$).
    Panels (c) and (d) show the trajectory shifts for a sharp step ($\Delta w = 0$)
    with $V_{0} = 0.7 \, t$ and $2.0 \, t$, respectively, and
    different wave packet sizes $\sigma_{k}$;
    here the wave packet energy is $E = 0.4 \, t$ ($V_{c} = 1.4 \, t$).
    In (d) the $\sigma_{k} = 0.05/a$ curve has been reduced by a factor of $0.5$.
    \label{fig:TAndRShifts}}
\end{figure}

To quantify the lateral shifts of the outgoing (asymptotic) trajectories
due to the anomalous velocity term in the semiclassical equation of
motions and to clarify that these shifts arise 
dominantly from the Berry curvature 
$\hat{\Omega}_{{\bf k}}$, we plot the differences of the
shifts occurring for positive and negative effective mass
as a function of $|\Delta \mu|$ in Fig. \ref{fig:TAndRShifts}. 
The sign of the Berry curvature, which depends on the sign of $\Delta \mu$
does not affect the energy dispersion.
Hence incoming and outgoing wave packet energies, momenta and group
velocities remain unaltered under such a sign change.
However, the topological properties of the Bloch wave functions change,
and therefore any accompanying changes in the wave packet trajectories
[see Fig. \ref{fig:trajectories} (a)]
are due to a change of $\hat{\Omega}_{{\bf k}}$.
The Berry curvature for electron and hole bands of $H_{0}$ 
can be determined explicitly \cite{Diener03arXiv}
and simplifies near the Dirac points ($|{\bf q}| \ll |{\bf K}|, |{\bf K}^{'}| $) to 
\cite{Dudarev04PRL}
\begin{eqnarray}
  \hat{\Omega}^{\nu}_{{\bf K}+{\bf q}} &\approx&
\nu \frac{9 a^{2} t^{2} \Delta \mu}{[4 \Delta \mu^{2} + 9
  |{\bf q}|^{2} a^{2} t^{2}]^{3/2}} \, \hat{\bf z}, \\
\hat{\Omega}^{\nu}_{{\bf K}^{'}+{\bf q}} &=& -  \hat{\Omega}^{\nu}_{{\bf K}+{\bf q}},
\end{eqnarray} 
where $\nu = \pm$ stands for the band index.

The lateral shifts in Fig. \ref{fig:TAndRShifts}
are defined as the projected shifts parallel to the armchair step
edge
\begin{eqnarray}
  \Delta s_{\text{tr}}(|\Delta \mu|) &=& {\bf s}_{\text{re}}(|\Delta \mu|) -
  {\bf s}_{re}(-|\Delta \mu|), \\
  \Delta s_{\text{re}}(|\Delta \mu|) &=& {\bf s}_{\text{tr}}(|\Delta \mu|) - {\bf s}_{\text{tr}}(-|\Delta \mu|),
\end{eqnarray}  
where ${\bf s}_{\text{re/tr}}(\Delta \mu)$ are the intersections 
of the corresponding outgoing asymptotes with the step edge.
(For wide steps we take the intersections between the 
asymptotes and the centre line of the step.)
Since the outgoing trajectory asymptotes for opposite 
staggered sublattice potentials ($\pm |\Delta \mu|$)
are roughly parallel, 
$\Delta s_{\text{tr}}(|\Delta \mu|)$ and $\Delta s_{\text{re}}(|\Delta \mu|)$ 
are well defined and can be interpreted as shown in Fig. \ref{fig:trajectories} (a).

Figure \ref{fig:TAndRShifts} displays the shifts of
wave packets coming in at normal incidence ($\alpha = 0 $),
with reflected and transmitted portions going out on likewise normal but
shifted trajectories.
The lateral displacements of the outgoing paths clearly reflect the
two transmission regimes discussed above.
Let us focus first on the left-hand side panels, where the wave packet
energy is $E = 0.2 \; t$, and the wave packet size is fixed at $\sigma_{k} = 0.1/a$.
For low step heights $V_{0} \lesssim V_{c}$ 
[Fig \ref{fig:TAndRShifts} (a)], 
where intravalley transmission is favoured, 
the lateral shifts of the transmitted trajectories (solid lines) are small compared
to those of the reflected trajectories (dashed lines) for
both sharp (black dots) and smooth (red triangles) potential steps.
For larger step heights $V_{0} > V_{c}$  [panel (b)], where
intervalley reflection dominates, the behaviour of the 
trajectory shifts reverses, with the largest lateral displacements 
occurring for the transmitted wave packet portion.
In the present cases the largest shifts are $\sim 40 a$, which is of
the order of the real space wave packet diameter at impact onto the
step edge ($ \sim 2 \, \sigma_{k}^{-1} = 20 \, a$). 

Panels (a) and (b) further suggest that 
steepening the potential steps strongly reduces the
lateral shifts of the reflected paths.
This finding goes beyond the semiclassical analytical approximation,
which relies on smooth potentials \cite{Chang96PRB,Xiao10RMP} and upon
integration of Eq. (\ref{eq:SemiClassEQM}) predicts that
\begin{eqnarray}
  \label{eq:lateralshift}
  \Delta s_{\text{re}} \approx - \int_{{\bf k}_{\text{in}}}^{{\bf k}_{\text{re}}} \text{d}{\bf k} \times
  \hat{\Omega}^{+}_{{\bf k}}
\end{eqnarray} 
is independent of the underlying potential shape
(${\bf k}_{\text{in}}$ and ${\bf k}_{\text{re}}$ are the central
momenta of the initial wave packet and the reflected wave packet
portion, respectively). \cite{Diener03arXiv}
Nonetheless, Eq. (\ref{eq:lateralshift}) further suggests that in the
intravalley regime $V_{0} \lesssim V_{c}$ the reflected (transmitted) shifts $\Delta
s_{\text{re}}$ ($\Delta s_{\text{tr}}$) are large (small), due to
constructive (destructive) integration of the Berry curvature within
the same band (in different bands).
The opposite is true for the intervalley regime $V_{0} \gtrsim V_{c}$
such that $\Delta s_{\text{tr}} > \Delta s_{\text{re}}$. 
The shifts reflect the topological
properties of the Bloch states, which can be 
manipulated valley-selectively
in graphene-like materials such as silicene 
by applying a perpendicular electric field and/or
photo-irradiation. \cite{Liu11PRB,Ezawa13PRL}

It is also interesting to observe that 
the reflected shifts are maximized for moderate $\Delta \mu$.
The maximum shifts are especially pronounced for
the reflected (transmitted) trajectories in panel (a) [(b)].
Furthermore, extrapolating to $\Delta \mu = 0$ all shifts seem to
disappear.
This suggests that higher order corrections (such as the second order term
discussed in Ref. \onlinecite{Gao14arXiv}) 
due to a sharp, discontinuous potential
are absent for $\Delta \mu = 0$
in spite of the broken inversion symmetry 
originating from the armchair step edge.

To study the effect of the wave packet size on the lateral shifts, 
we have considered
wave packets at normal incidence and
at a slightly higher energy $E = 0.4 \, t$.
A higher energy moves the initial central momentum further 
away from the Dirac point K'. 
This allows larger variation of the momentum space wave packet size
$\sigma_{k}$ without creating highly dispersive real space wave
packets that appear when K' is included in the 
momentum space support. \cite{footnote01}
Here we also focus on sharp steps only ($\Delta w = 0$).
The resulting lateral shifts are displayed in
Fig. \ref{fig:TAndRShifts} (c) and (d) representing
both Klein tunneling regimes $V_{0} \lesssim V_{c}$ and $\gtrsim
V_{c}$ with $V_{c} = 1.4 \, t$, respectively.
The overall behaviour is similar to the low energy wave packet
shifts at $E = 0.2 \, t$ for sharp steps [cf. black curves with dots in (a) and (b)].
While the transmitted (reflected) shifts are largely unaffected 
by variation of $\sigma_{k}$ for $V_{\text{0}} \lesssim V_{c}$ ($\gtrsim V_{c}$),
the reflected (transmitted) shifts increase with decreasing 
momentum wave packet size $\sigma_{k}$ 
(i.e. with increasing initial real space wave packet size).
This indicates that semiclassical approximations of 
wave packet shifts like Eq. (\ref{eq:lateralshift}) are corrected and
significantly weighted by the overall wave packet 
and Berry curvature distributions in momentum space.

\section{Valley Hall effect and Bloch oscillations}
\label{sec:IV}
A sharp or smooth step imposes an electric field locally.
In the opposite limit of a uniform electric field, 
Berry curvature induced shifts can also be observed in the
wave packet motion. \cite{Diener03arXiv}
In the presence of a uniform electric field along the $\Gamma$-K
direction, electron wave packets 
perform Bloch oscillations in real space and periodically pass
through K and K' valleys in momentum space.
Near K and K' interband Bloch-Zener transitions redistribute the
wave packet weights in both bands such that in real space an initially
one component wave packet splits up into
two opposite oscillating wave packets.
By including a finite Berry curvature, the trajectories are laterally shifted at each
pass through a valley.
This is the semiclassical manifestation of the valley Hall effect. 
\cite{Diener03arXiv,Xiao10RMP}

\begin{figure}[t!]
  \centering
  \includegraphics*[width=1.0\linewidth, clip]{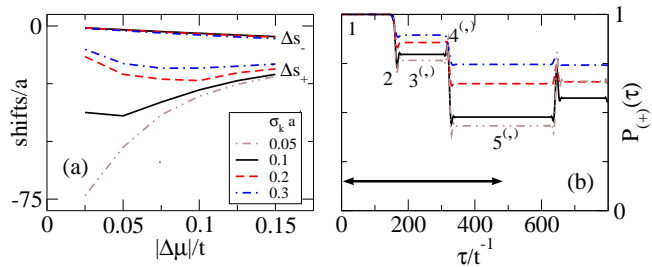}
  \caption{(Color online) 
     Lateral shifts and interband Bloch-Zener transitions
    associated with Bloch oscillations / valley Hall effect
    for the parameters used in Fig. \ref{fig:BO}.
    (a) Lateral shifts for various (initial) wave packet sizes $\sigma_{k}$
    as a function of staggered sublattice potential
    $|\Delta \mu|$; the shifts are indicated in Fig. \ref{fig:BO}.
    (b) Electron band weight of the oscillating 
    wave packets as a function of time $\tau$. 
    The jumps correspond to Bloch-Zener transitions at the Dirac points.
    The black double arrow indicates one Bloch oscillation period.
    The numbers correspond to the temporal sequence displayed in Fig. \ref{fig:BO}.
    Please see main text for details.
    \label{fig:BO2}}
\end{figure}

The resulting trajectories 
are shown in Fig. \ref{fig:BO} for a uniform electric field of magnitude
$F = 0.015 t/(|e| \, a)$ oriented normally to the armchair direction.
Here the initial Gaussian-like wave packet 
is centred at the $\Gamma$ point (${\bf k}_{0} = {\bf 0}$)
with full weight of the wave function given to the electron band.
\cite{footnote03}
The real space and momentum space positions of the wave packet 
at intermediate points during one oscillation period are 
indicated by the numbers in Fig. \ref{fig:BO} for the red solid ($\mu =
0.1 \, t$)  trajectory pair. 
Panel (b) of Fig. \ref{fig:BO2} displays the corresponding temporal evolution of the
wave packet's electron band weight
\begin{eqnarray}
  P_{(+)}(\tau) = \sum_{{\bf k}} \rho^{(+)}({\bf k}; \tau),
\end{eqnarray}
with the numbers corresponding to those
in Fig. \ref{fig:BO}.
During one oscillation period (sequence $1 \rightarrow 5^{(\prime)}$) 
the wave packet is driven through the K' and the K valley. 
Each time the wave packet passes through a valley (points 2 and 4),
a portion undergoes a Bloch-Zener transition from the electron band to
the hole band [and vice versa], 
splitting the wave packet in real space due to the opposite group velocities.
The interband portion's velocity in real space remains roughly unchanged
(i.e., continues from point 2 of the lower red solid trajectory to point $3^{\prime}$
of the upper red solid trajectory in Fig. \ref{fig:BO}),
while the intraband part's velocity reverses its sign (i.e., propagates
from point 2 to 3 on the lower red solid trajectory).
During the valley passage, the intraband portion of the wave packet
experiences a Berry curvature induced side shift.
The shift of the interband portion on the other hand is minute.
Both types of lateral shifts are indicated in Fig. \ref{fig:BO}, where 
$\Delta s_{+}$ ($\Delta s_{-}$)
stands for the intraband (interband) valley Hall shift, respectively, 
and quantified in Fig. \ref{fig:BO2} (a) for various initial wave packet sizes
$\sigma_{k}$.
The behavior of the shifts $\Delta s_{+}$
and $\Delta s_{-}$ is in agreement with that of 
$\Delta s_{\text{re}}$ and $\Delta s_{\text{tr}}$ 
in the case of the step potential, respectively.
Interestingly, as before, the shifts are largest for small $\sigma_{k}$ (i.e.,
for large initial real space wave packet sizes). 
In the limit of an extended Bloch wave, trajectories and shifts are
not defined but a phase slip in the wave function portions
may be expected due the finite Berry curvature.  
We also note, that, as implied by Eq. (\ref{eq:lateralshift}) and
despite the $\sigma_{k}$-dependence, the shifts are
largely independent of the field strength $|{\bf F}|$.
 
The Bloch-Zener transitions causing the jumps in Fig. \ref{fig:BO2} (b)
are suppressed for large gaps and for small driving forces.
For these reasons the jumps at the first two transitions (at $\tau
\approx 160/t$ and $320/t$) are largest for wave
packets with small $\sigma_{k}$
as most of the wave packet experiences only a small gap at
the Dirac point.
At later jumps the wave packet has contributions and transitions
from and to both bands, so that after long times 
one expects $P_{+} (\tau) = 0.5$.

\section{Summary and conclusion}
We have investigated in detail the semiclassical electron dynamics in 
graphene using a numerical wave packet simulation.
Compared to the semiclassical analytical approach
\cite{Xiao10RMP,Gao14arXiv} 
the numerical simulations are more general and in principle exact.
They are based on the full graphene tight-binding Hamiltonian and can
account for intervalley, interband and Berry phase related phenomena.

The present simulations establish both intra- and intervalley Klein
tunneling processes.
The latter becomes relevant for sufficiently tall potential steps
and has received only limited attention in graphene-based
Klein tunneling studies.
In this limit, intervalley transmission is greatly enhanced in the presence
of a finite gap and/or by smoothing the step profile
due to the lack of pseudospin conservation.
The presence of a gap furthermore leads to notable Berry curvature induced trajectory
shifts, which for the cases considered, are roughly of 
the order of the wave packet size.
Such trajectory modifications are reminiscent of the Hall effect
of light in three dimensions,
where an electromagnetic wave packet scatters at an interface between two
optically different media. 
Here reflected and transmitted light rays are similarly laterally shifted at the
interface due to Berry curvature corrections. \cite{Onoda04PRL,Onoda06PRB}

Applying a linear potential, our simulations illustrate
the valley Hall effect in form of laterally shifted 
Bloch oscillation trajectories. \cite{Diener03arXiv}
Accompanying Bloch-Zener transitions redistribute the wave
function among the graphene bands, hence giving rise to two 
wave function portions with opposite oscillating motion in real space.

The present study highlights valley-dominated scattering dynamics of
electrons in a honeycomb lattice.
The energy bands and valley properties of silicene, a close relative
of graphene, have been suggested to be easily manipulable experimentally; in fact,
they may be tuned to exhibit a variety of topological
properties, ranging from single Dirac-cone states to various types of 
quantum Hall insulators. \cite{Ezawa13PRL}
Such manipulations would have direct impact on the intra- and
intervalley wave packet dynamics discussed here.   

The simulation results also make explicit the dependence of the trajectory
shifts on the wave packet size and the effective mass
(i.e. the Berry curvature distribution in momentum space).
Reducing the initial momentum space wave packet size ($\sigma_{k}$) can
drastically enhance the trajectory shifts in both regimes.  
Furthermore, for a fixed wave packet size, there is an optimum effective mass
that maximizes the shifts of the outgoing trajectories 
and the widths of the Bloch oscillation paths.

Experimentally, the trajectory shifts may be observed as scattering
asymmetry at np junctions \cite{Stander09PRL,Sajjad12PRB,Rahman13arXiv} 
for either reflected or transmitted
electrons, where a staggered sublattice potential may originate from an  
appropriate substrate material or, in the case of silicene and
germanene, from an external perpendicular electric field. \cite{Liu11PRB,Ezawa13PRL}
In the limit of a single spatially extended Bloch state, Berry
curvature induced shifts become phase jumps
at the step interface, which may be observed as 
anomalies in the conductance oscillations across a potential
barrier due to interference between forward and backward reflected 
wave functions. \cite{Young09NatPhys}
These shifts may also substantially affect the electron focusing and lensing
behaviour of potential barriers and potentials steps
in electron optics applications.
\cite{Cheianov07Science}

It would further be interesting to investigate how the 
wave packet dynamics change for zigzag and Klein edges.
The dynamics at zigzag and Klein edges are 
less symmetric about the interface normal than in the armchair case.
For sharp steps this  will affect the direction and the Berry
curvature induced shifts of the wave packet trajectories.
The coupling of wave packets with edge states
may give rise to further intriguing behavior
at resonance energies.

\vspace{0.4cm}

The authors thank T. Nakanishi, K. Wakabayashi, N. Kumada,
O. Entin-Wohlman and A. Aharony for useful discussions.
Y. H. acknowledges partial financial support through
 Grants-in-Aid (\mbox{KAKENHI} No. 23340112 and No. 23654128) from
 JSPS.
Part of this work is supported financially by the Funding Program for
World-Leading Innovative R\&D Science and Technology (FIRST).

\end{document}